\documentstyle[12pt,aasms]{article}

\begin{document}

\title{Solar Activity and Cloud Opacity Variations:
A Modulated Cosmic-Ray Ionization Model}

\author{David Marsden}
\affil{Scripps Institution of Oceanography\\ University of 
California, San Diego\\ 9500 Gilman Dr., Dept. 0242 \\ La Jolla, 
California 92093-0242 \\ email: dmarsden@ucsd.edu}

\author{Richard E. Lingenfelter}
\affil{Center for Astrophysics and Space Sciences\\
University of California, San Diego}

\slugcomment{8/23/02}

\begin{abstract}

The observed correlation between global low cloud amount and
the flux of high energy cosmic rays supports the idea that
ionization plays a crucial role in tropospheric cloud formation. 
We explore this idea quantitatively with a simple model linking 
the concentration of cloud condensation nuclei to the varying 
ionization rate due to cosmic rays. Among the predictions of the 
model is a variation in global cloud optical thickness, or opacity, 
with cosmic-ray rate. Using the International Satellite Cloud 
Climatology Project database (1983--1999), we search for variations 
in the yearly mean visible cloud opacity and visible cloud amount due 
to cosmic rays. After separating out temporal variations in the data 
due to the Mt. Pinatubo eruption and El Ni\~{n}o/Southern Oscillation,
we identify systematic variations in opacity and cloud amount due to 
cosmic rays. We find that the fractional amplitude of the opacity variations 
due to cosmic rays increases with cloud altitude, becoming approximately 
zero or negative (inverse correlation) for low clouds. Conversely, the 
fractional changes in visible cloud amount due to cosmic rays are only 
positively-correlated for low clouds and become negative or zero for 
the higher clouds. The opacity trends suggest behavior contrary to the 
current predictions of ion-mediated nucleation (IMN) models, but more 
accurate temporal modeling of the ISCCP data is needed before 
definitive conclusions can be drawn.

\end{abstract}

\section{Introduction}
\label{intro}

The primary source of energy for the Earth's atmosphere is the 
Sun, so it is reasonable to explore whether changes in the global 
climate result from solar variability. It was first suggested by 
the astronomer William Herschel (\cite{herschel01}) that variations 
in the solar irradiance caused by sunspots could lead to climatic 
changes on Earth, and he cited the variation of British wheat prices 
with sunspot number as evidence for this link. The occurrence of the 
``Little Ice Age'' during the 1645-1715 Maunder sunspot minimum 
(\cite{eddy76}), the correlation between the long-term solar cycle 
variations and tropical sea surface temperatures (\cite{reid87}), 
polar stratospheric temperatures (\cite{labitzke87}), and the width 
of tree rings (\cite{zhou98}), along with many other studies also 
support a link between solar variations and the Earth's climate.

A direct link between the Sun and these phenomena is tenuous,
however, because the magnitude of the solar irradiance variation
over the 11-year solar cycle is very small. Over the 1979-1990 solar
cycle, for example, the variation in the irradiance was only $\sim
0.1\%$ (\cite{frohlich00}), or $\sim 0.3$ W m$^{-2}$ globally-averaged 
at the top of the atmosphere. This is insufficient to power the sea 
surface temperature changes associated with the solar cycle by a factor 
of $3-5$ (\cite{lean97}), and is significantly smaller than the 
globally-averaged forcings due to clouds ($\sim 28$ W m$^{-2}$; e.g. 
\cite{hartmann93}), anthropogenic greenhouse gases ($\sim 2$ W m$^{-2}$; 
\cite{wigley92}), and anthropogenic aerosols ($\sim 0.3-2.0$ W m$^{-2}$; 
\cite{charlson92}; \cite{kiehl93}), suggesting that any direct atmospheric 
forcing from solar irradiance variations would be relatively unimportant.

An indirect link between solar cycle variations and the Earth's
climate appears more likely, especially given the discovery of
a link between the flux of Galactic cosmic rays (GCRs) and global 
cloudiness (\cite{svensmark97}) in the ISCCP cloud database 
(\cite{rossow99}). The Sun modulates the GCR flux at the Earth
through the action of the solar wind, which scatters and attenuates
the GCRs in times of heightened solar activity (solar maximum; e.g. 
\cite{jokipii71}). Using $3.7$ $\mu$m infrared (IR) cloud amounts 
from the ISCCP database for the years 1983-1993, Marsh and Svensmark 
(2000) and Pall\'{e} Bag\'{o} and Butler (2000) showed that there is 
evidence of a positive GCR-cloud correlation only for low ($<3$ km) 
clouds, and that the effect of the cosmic rays on global cloud amount 
appears to be greatest at the low to mid latitudes. The globally-averaged 
forcing due to the increase in low clouds associated with the solar cycle 
GCR variations is estimated (\cite{kirkby00}) to be approximately $-1.2$ 
W m$^{-2}$, which is sufficient to power the sea surface temperature 
variations (\cite{lean97}). This is also comparable in magnitude (but 
opposite in sign) to the forcing due to anthropogenic CO$_{2}$ emission 
over the last century (\cite{svensmark97}). Decreasing local cloud amounts 
correlated with short-term Forbush decreases in cosmic-ray rates were 
observed by Pudovkin and Veretenko (1995).

The reality of the GCR-cloud connection has been questioned by 
a number of authors (\cite{kernthaler99}; \cite{jorgensen00}; 
\cite{norris00}). These objection can be distilled into three
main points: 1) The GCR-cloud correlation should be seen prominently
in high (cirrus) clouds at high latitudes where the cosmic-ray 
intensity is highest, 2) the increased cloudiness can be more 
plausibly attributed to other phenomena instead of GCRs, and 3) 
the correlation is an artifact of the ISCCP analysis. The first 
objection is addressed by the theory of ion-mediated nucleation 
(IMN: \cite{yu01}; Yu 2002), in which the efficiency of the cosmic-ray 
interaction is limited at high altitudes by the lack of aerosol 
precursor vapors such as H$_{2}$SO$_{4}$ relative to the ion 
concentration. For the second objection, the temporal profile 
of the GCR-cloud correlation may be inconsistent with the profiles 
of the dominant volcanic and El Ni\~{n}o/Southern Oscillation (ENSO) 
events during the same time period (\cite{kirkby00}), although no 
quantitative study of the various temporal signatures in the data 
has been undertaken. Finally, the ISCCP artifacts pointed out by Norris 
(2000) are troubling, but it is not clear that they are of sufficient 
magnitude to produce the observed GCR-cloud correlation, and it doesn't 
explain why the correlation exists only for low clouds and not the other 
cloud types in the ISCCP database.

The linkage between cosmic rays and cloud formation has been recently
investigated by a number of authors (\cite{yu02}; \cite{yu01}; 
\cite{tinsley00} and references therein). Here we apply a perturbative 
approach to quantify the effects of variations in the cosmic-ray rate 
on the optical thicknesses, or opacities, of clouds, and use the observed 
cloud opacity variations to constrain the microphysical models 
of ion-mediated ultrafine particle formation. The paper is organized as 
follows. In the next section we discuss how the effect of cosmic rays 
could alter the optical thickness and emissivity of clouds by affecting 
the nucleation of condensation nuclei (CN). The search for variations in 
cloud optical properties using the ISCCP database and their correlation 
with cosmic-ray flux variations are discussed in Section \ref{variations} 
A discussion of the results is given in Section \ref{discussion}, and 
finally we summarize our results in Section \ref{summary}

\section{Effects of GCRs on Cloud Properties}
\label{effects}

\subsection{Nucleation}
\label{nucleation}

Cosmic rays form water droplets in the supersaturated air of a
classical cloud chamber (\cite{wilson01}), and it seems plausible 
that they could also play a significant role in natural cloud 
formation. Yu and Turco (2000, 2001) and Yu (2002) have investigated 
the formation of ultrafine CN from charged molecular clusters 
formed from cosmic-ray ionization, and they find that the charged 
clusters grow more rapidly and are more stable than their neutral 
counterparts up to a size of $\sim 10$ nm. Although the subsequent 
growth of the cosmic-ray formed ultrafine CN to viable $\sim 100$ nm 
cloud condensation nuclei (CCN) has not been explored, the concentration 
of CCN should also reflect the CN concentration, as well as the direct 
influence of cosmic rays, if the cosmic-ray ionization rate does not 
affect other important nucleation efficiency parameters such as 
condensible vapor concentration, temperature, and pressure. We 
will make this assumption here although it may not be strictly 
true with respect to the condensible vapor concentration (see 
e.g. \cite{turco00}; \cite{yu02}).

Althought the formation of CCN and ultimately cloud droplets 
is a function of many variable factors such as temperature, 
pressure, vapor concentration, and relative humidity, we can 
quantify the effects of small variations in the ionization 
rate (primarily due to cosmic rays above ocean and at altitudes 
$>1$ km above land; e.g. \cite{reiter92}) on the number of CCN through 
a perturbation approach, i.e. 
\begin{equation}
\label{perturb}
N_{\rm{CCN}}(q+\Delta q,V)\approx N_{\rm{CCN}}(q,V)+\Delta q
\left .{\partial N_{\rm{CCN}}\over \partial q}\right |_{V},
\end{equation}
where $N_{\rm{CCN}}$ is the concentration of CCN, $q$ is the 
ionization rate, $V$ refers to the set of parameters other than the 
ionization rate affecting $N_{\rm{CCN}}$, and the partial derivative 
is evaluated for fixed $V$ (hereafter this will not be written 
explicitly). Along with the assumption discussed previously, this 
approach assumes that the quantity $\Delta q|\partial N_{\rm{CCN}}/
\partial q|<<N_{\rm{CCN}}(q,V)$, which is probably true for solar 
cycle variations, where $q$ typically varies by $<30\%$, but may not 
be true during periods of large scale changes in the geomagnetic 
field (e.g. \cite{tric92}).

To quantify the effect of varying CCN concentrations on cloud optical 
thicknesses, we envision the two idealized scenarios depicted in 
Figure~1. In both cloud formation scenarios, changes in the 
ionizing cosmic-ray flux cause changes in the number of cloud 
condensation nuclei through the process of ion-mediated nucleation 
on the formation of ultrafine CN in accordance with the assumptions 
mentioned above\footnote{We have assumed $\partial N_{\rm{CCN}}/
\partial q>0$ in Figure~1, which need not be valid for all $q$.}. In 
the first case we assume that the nucleation of cloud droplets is 
limited by the available amount of water in the supersaturated air, 
so that the liquid water content (LWC), or density of water in droplets, 
is constant. Therefore the amount of water per droplet and hence the 
effective radii of cloud droplets will change with the cosmic-ray 
ionization rate. This is analogous to the ``Twomey Effect'' of 
enhanced aerosol pollution on droplet size distributions and the 
albedo of clouds (\cite{twomey77}; \cite{rosenfeld00}), and would 
primarily occur in environments where the amount of water in the air 
(and not the number of CCN) is the limiting factor. Thus, using 
(\ref{perturb}), we would expect that the effective radius $R_{eff}$ 
of the cloud droplet distribution resulting from a small change in 
the cosmic-ray ionization rate $\Delta q$ in any particular volume of 
air will be
\begin{equation}
\label{reff}
R_{eff}={\left [N_{\rm{CCN}}(q,V)\over N_{\rm{CCN}}(q+\Delta 
q,V)\right ]}^{1/3}R_{eff}^{0}\approx{\left (1 + {\Delta q\over 
N_{\rm{CCN}}}{\partial N_{\rm{CCN}}\over \partial q}\right )}^{-1/3} 
R_{eff}^{0},
\end{equation}
where $R_{eff}^{0}$ is the effective radius of the unperturbed droplet 
distribution, which we will associate with the solar maximum period of 
the solar cycle.

In the second case in Figure~1, we assume that the change in CCN 
concentration resulting from change in cosmic-ray ionization causes 
a proportionate change in the amount of water extracted from the 
supersaturated air, with the effective radius of the cloud droplet 
distribution remaining constant. This is the case where the formation 
of the cloud is limited by the local availability of CCN and not 
condensible water. This effect has been seen in the marine boundary 
layer in ship track clouds (\cite{conover66}), which have higher 
reflectivities (\cite{coakley87}) and liquid water contents 
(\cite{radke89}) due to the formation of additional ultrafine 
CN from ship exhaust. The perturbed liquid water content of a 
cloud in any particular volume of air will then be given by
\begin{equation}
\label{lwc}
\rm{LWC} \approx \left (1 + {\Delta q\over N_{\rm{CCN}}}{\partial 
N_{\rm{CCN}}\over \partial q}\right )\rm{LWC}_{0},
\end{equation}
where $\rm{LWC}_{0}$ is the unperturbed cloud liquid water content 
associated with solar maximum as before. These two scenarios probably 
represent extremes of the direct cosmic-ray ionization effect on the 
clouds. As in the ship track clouds, the effect of the GCRs will 
probably be a combination of both LWC changes and $R_{eff}$ changes, 
with the magnitude of the effect being bounded by the changes given 
in (\ref{reff}) and (\ref{lwc}).

\subsection{Radiative Properties}
\label{optical}

Changes in the cloud liquid water content and droplet effective radius, 
associated with changes in the ionization rate due to cosmic rays, will 
result in changes in cloud opacities. The optical thickness $\tau$ of a 
uniform cloud layer of thickness $\Delta z$ is given by (\cite{vandenhulst81}):
\begin{equation}
\label{tau}
\tau=\Delta z \int_{0}^{\infty}Q_{ext}\,n(r)\pi r^{2}dr,
\end{equation}
where $n(r)dr$ is the concentration of cloud droplets with radii 
between $r$ and $r+dr$, $Q_{ext}$ is the Mie extinction efficiency,
and it is commonly assumed that
\begin{equation}
\label{int}
{\int_{0}^{\infty}Q_{ext}n(r)r^{2}dr\over \int_{0}^{\infty}n(r)
r^{2}dr}=2,
\end{equation}
which is a good approximation when $2\pi r/\lambda>>1$,
where $\lambda$ is the wavelength (\cite{stephens84}).

The effective radius of the cloud droplet distribution is given by
\begin{equation}
\label{reff2}
R_{eff}={\int_{0}^{\infty}n(r)r^{3}dr\over \int_{0}^{\infty}n(r)
r^{2}dr},
\end{equation}
and the cloud liquid water content is given by
\begin{equation}
\label{lwc2}
\rm{LWC}={4\over 3}\pi \rho\int_{0}^{\infty}{n(r)r^{3}dr},
\end{equation}
where $\rho$ is the density of liquid water. Combining these 
equations, we see that
\begin{equation}
\label{tau2}
\tau\approx {3\over 2}{{\rm LWC}\,\Delta z\over \rho\, R_{eff}}.
\end{equation}
Thus from (\ref{tau2}) we would expect that an increase (decrease) in 
the mean $R_{eff}$ and a decrease (increase) in the mean $\rm{LWC}$, 
resulting from ionization variations due to cosmic rays, would result 
in a decrease (increase) the mean opacity of clouds. 

The change in cloud opacity with cosmic-ray rate can be quantified using 
the perturbation assumptions discussed in Section \ref{nucleation} and 
equations (\ref{reff}), (\ref{lwc}), and (\ref{tau2}). The fractional 
change in cloud opacity is then given by
\begin{equation}
\label{dtaureff}
{\delta\tau\over \tau} \sim {\Delta q\over fN_{\rm{CCN}}}{\partial 
N_{\rm{CCN}}\over \partial q},
\end{equation}
where $f=1(3)$ for CCN (water) limited cloud formation, and the fractional 
change in the perturbed opacity $\tau$ (relative to the unperturbed opacity 
$\tau_{0}$) is defined by $\delta\tau/\tau=(\tau - \tau_{0})/\tau_{0}$. 
As mentioned previously, this derivation assumes that the right hand side 
of (\ref{dtaureff}) is much less than one, which may not be the case for 
large changes in $q$ and $N_{\rm{CCN}}$. As before we will assume that 
the unperturbed (perturbed) values of $q$ and $N_{\rm{CCN}}$ refer to 
the values at solar maximum (minimum).

At visible wavelengths from space, the primary consequence of the change 
in cloud opacity associated with cosmic rays will be an increase in 
cloud reflectivity, or albedo. To investigate this, we use the 
radiative transfer code SBDART (\cite{ricchiazzi98}) to calculate 
the top of the atmosphere broadband ($0.25-4.00$ $\mu\rm{m}$) upward 
flux for three uniform low cloud models: 1) a $1$ km thick cloud layer 
extending to a height of $2$ km, 2) a $2$ km thick cloud extending to 
a height of $3$ km, and 3) a $0.5$ km cloud layer extending to $1.5$ 
km. These simulations were done with a tropical atmosphere profile 
(\cite{mclatchey72}) and an ocean surface albedo. The fractional 
increases in albedo, resulting from a 10\% increase in the number of 
cloud droplets due to cosmic ray ionization variations, is shown in 
Figure~2 for the $1$ km thick cloud case, for a wide range of LWC and 
$R_{eff}$ in the variable LWC case (top panel) and the variable 
$R_{eff}$ case (bottom panel). In both cases the contours of changing 
albedo approximately parallel the change in optical thickness calculated 
assuming $Q_{ext}=2.0$ 

Figure~3 shows the fractional change in albedo directly as a function 
of opacity for all three cloud models. This figure clearly shows that 
the change in albedo is largest for clouds with opacities $\tau$ between 
1 and 10, but is roughly independent of cloud geometrical thickness. 
Figures~2 and 3 indicate that the change in cloud optical thickness 
can be used to quantify the effects of the cosmic rays on cloud optical 
properties. Although the fractional change in albedo due to the cosmic rays 
is only $\sim 2-5\%$ for a $10\%$ variation in the number of cloud droplets, 
this can produce a significant forcing per cloud of $\sim 7-16$ W m$^{-2}$ 
at the top of the atmosphere for a solar zenith angle of $40^{\circ}$. The 
modulation of cloud opacity due to cosmic rays could therefore produce a 
similar modulation of the Earth's energy budget over the 11 year solar cycle, 
although the exact amount of forcing due to cosmic rays will depend sensitively 
on cloud amount variations, cloud opacity variations, and the efficiency at 
which changes in the cosmic ray rate are reflected in the number of cloud 
condensation nuclei.

Because of the relationship between cloud opacity and emissivity, 
the cosmic rays should also produce an observable effect on cloud 
emission at infrared (IR) wavelengths. The effective IR emissivity 
$\epsilon$ can be parameterized by a relation of the form 
(\cite{stephens78}):
\begin{equation}
\label{emiss}
\epsilon=1 - \exp(-a_{0}\rm{LWC}\,\Delta z),
\end{equation} 
where $a_{0}$ is the mass absorption coefficient. Empirical fits to 
IR emission from water clouds yield $a_{0}=0.130$ (\cite{stephens78}). 
The exponent in (\ref{emiss}) is proportional to the cloud optical 
thickness for a given droplet effective radius, so the infrared 
emissivity increases with cloud opacity, with the change being most 
noticeable for optically thin clouds. Therefore one would expect a 
change in IR emission, along with the primary effect of changes in 
visible albedo, from clouds at solar minimum relative to clouds at 
solar maximum if the cosmic rays change the cloud liquid water 
contents. Interestingly, a correlation between cosmic ray rate 
and cloud top temperature for low clouds has been reported by 
Marsh and Svensmark (2000), supporting this hypothesis.

\section{Cloud Opacity Variations}
\label{variations}

\subsection{ISCCP Data}
\label{data}

To search for systematic temporal changes in synoptic scale cloud optical 
properties, we used the International Cloud Climatology Project (ISCCP)
monthly gridded cloud products (``D2'') datasets, a compilation of cloud
properties derived from satellite observations during the period 1983--1999 
(\cite{rossow99}). The ISCCP D2 data used here consists of mean daytime 
cloud amount fractions and visible optical depths, as a function of time, 
for $6596$ ``boxes'' with equal area covering the entire surface of the 
Earth. For a given time, the cloud amount fraction in each box is defined 
as the number of cloudy satellite image pixels, as determined by a cloud 
detection algorithm, divided by the total number of pixels in the 
box. The cloud optical thicknesses are derived from the visible 
satellite cloud albedos by using a radiative transfer model and assuming 
spherical droplets with droplet sizes characterized by a gamma distribution 
with variance $0.15$ and $R_{eff}=10$ $\mu$m. ISCCP cloud top temperatures 
are simultaneously determined from the $3.7 \mu\rm{m}$ IR radiances, 
allowing for determination of cloud altitude and pressure, and the low, 
mid-level, and high clouds are defined as having cloud top pressures 
$P>680$ mb, $440<P<680$ mb, and $P<440$ mb, respectively. Because we 
require the simultaneous visible and infrared radiances to determine 
the opacity and cloud height for our analysis, we only use the ISCCP 
daytime data. This is a different dataset than the diurnal 1983-1993 
IR data used for the cloud amount analyses of Marsh and Svensmark 
(2000) and Pall\'{e} Bag\'{o} and Butler (2000).

Detailed information on the distribution of cloud optical thicknesses 
is not preserved in the the ISCCP D2 database, and instead the mean 
optical thickness $\bar{\tau}_{i}$ is recorded for three broad opacity 
bands $i$: $0.0-3.6$, $3.6-23.0$, and $23.0-379.0$. Thus a detailed analysis 
of the change in $\tau$ over the solar cycle is not possible using the 
D2 data, but a value of the {\it weighted} mean cloud optical thickness 
$\bar{\tau}$ can be calculated using
\begin{equation}
\label{taumean}
\bar{\tau}={\sum_{i=1}^{3}\bar{A}_{i}\bar{\tau}_{i}\over
\sum_{i=1}^{n}\bar{A}_{i}},
\end{equation}
where the $\bar{A}_{i}$ are the total mean cloud amount fractions within each 
of the broad ISCCP optical thickness bins mentioned above. We calculated
$\bar{\tau}$ separately for the three cloud altitude levels and for two
latitude bands with $|\phi|\leq 40.0^{\circ}$ (low latitude) and $|\phi|
>40^{\circ}$ (high latitude). The error associated with each $\bar{\tau}_{i}$ 
was estimated by calculating the standard deviation of each ISCCP data 
point, from the scatter about the mean, and scaling by the square root 
of the number of data points. 

The mean optical thicknesses $\bar{\tau}$ as a function of 
time for the low latitude clouds are shown in Figure~4, and the 
corresponding result for global high latitude clouds is shown in 
Figure~5. Shaded is the two year period in which the effects of the 
Mt. Pinatubo eruption appear to be most significant. Also shown for 
comparison are the mean counting rates from the Climax, Colorado 
neutron monitor run by the University of Chicago (obtained from 
http:$//$ulysses.uchicago.edu$/$NeutronMonitor$/$neutron\_mon.html), 
which is a good measure of the local cosmic-ray ionization rate. In 
the low latitude case, the abrupt and large decrease in $\bar{\tau}$ 
during 1991--1993 is due to the eruption of Mt. Pinatubo, and the 
subsequent plot scaling obscures smaller scale opacity variations. 
For comparison, we also plot the total mean cloud amount fractions 
$A=\sum_{i=1}^{3}\bar{A}_{i}$ for the same two latitude bands in 
Figures~6 and 7. These plots show evidence for increases in 
mean cloud amount due to Mt. Pinatubo, as well as the smaller-scale 
temporal variations. 

\subsection{Extracting The Cloud Variations Due to Cosmic-Rays}
\label{model}
	
To search for subtle variations in the ISCCP cloud opacities and amounts 
due to cosmic rays only, it is first necessary to eliminate the opacity 
variations in the data due to the Mt. Pinatubo volcanic eruption in 
June-September 1991 and strong ENSO events during the period of the 
ISCCP data. To separate out the various temporal signatures in the 
ISCCP data, we use a linear temporal model of the form
\begin{equation}
\label{tmodel}
F(t)=\sum_{k=0}^{3}b_{k}X_{k}(t),
\end{equation} 
where $F(t)$ is the mean ISCCP quantity of interest for the year $t$, 
which for our purpose is either the visible cloud opacity $\bar{\tau}$ 
or the mean cloud amount/fractional area $A$. The model consists of four 
temporal basis vectors $X_{k}$, which are functions of time, each 
scaled by a linear coefficient $b_{k}$. For our temporal model we 
choose basis vectors corresponding to constant level of the given 
quantity ($k=0$) and variations due to ENSO events (e.g. \cite{kuang98}), 
the Mt. Pinatubo eruption of 1991, and cosmic rays ($k=1$-$3$, respectively). 
Given the functional form of the basis vectors, the best-fit values of the 
linear coefficients can be determined through least squares minimization, 
and the fractional change in the time-varying ISCCP quantity over the data 
stretch is then given by $\delta F/F=b_{k}/b_{0}$, where $k=1$-$3$. This 
model assumes a linear correlation between the quantity of interest and 
the basis vectors and assumes no time delays; more complicated models are 
possible but will not be considered here. 

The normalized basis vectors used in the temporal analysis of the ISCCP 
cloud data are shown in Figure~8. All of the vectors are scaled to values 
between zero and one. For the ENSO term $X_{1}$ we use the scaled 
yearly-averaged Southern Oscillation Index (SOI) from the Australian 
Bureau of Meteorology (obtained from 
http://www.bom.gov.au/climate/current/soihtm1.shtml). The SOI is a measure 
of the size of fluctuations in the sea level pressure difference between 
Tahiti and Darwin, Australia, and small values of the scaled SOI denote 
El Ni\~{n}o conditions and large values La Ni\~{n}a -- both of which affect 
global weather (\cite{rasmusson82}). To parameterize the effect of the Mt. 
Pinatubo eruptions of 1991, we adopt a simple step function for $X_{2}$, 
with identical non-zero intensities only for years 1991 and 1992. For the 
final term in the temporal model, $X_{3}$, we use the scaled cosmic-ray rate 
from the Climax neutron monitor. Neutron monitor rates are directly 
proportional to the ionization rates due to cosmic rays because the 
neutrons are produced by the same cosmic ray cascade particles that 
produce the ionization, and the neutrons subsequently diffuse through 
less than 100 m of air before they are thermalized and captured by N to 
form $^{14}$C (e.g. Lingenfelter, 1963). Neutron counters are thus 
unsusceptible to background ionizations due to terrestrial radiation 
from radioactive decays, which dominate the ionization signal from 
Galactic cosmic rays only below $\sim 1$ km in the atmosphere 
(\cite{reiter92}).

The results of the temporal fitting of both the ISCCP visible cloud 
opacities and amounts are shown in Table~1. Formally most of the 
fits are not good, with reduced chi-squares ranging from $\sim 0.7-7.8$ 
for twelve degrees of freedom. There are a number of possible factors 
that could be contributing to this. For example the error bars on the 
data may have been underestimated, leading to an artificially large 
values of chi-squared. Another possibility is that our fitting model 
is missing other significant temporal drivers, or perhaps a non-linear 
model or different basis vectors may be required to fit the data. 
We tried to fit the ISCCP data with linear models composed of different 
combinations of our four basis vectors, and models with the cosmic-ray 
term provided a better fit to the data in general. Nevertheless it is 
possible that un-modeled phenomena mimic the temporal signature of 
cosmic rays in the data; more robust calculations of ISCCP error bars, 
inclusion of more ISCCP data, and exploration of more complicated temporal 
models in future work will help resolve this issue. 

The fractional variation in visible opacity $\delta\tau/\tau$ associated 
with the cosmic rays ranges from $\sim +10\%$ for high clouds to $-7\%$ 
for low clouds. For the mean visible cloud amounts the variation due to 
cosmic rays is just the opposite -- becoming greater in magnitude as the 
cloud height {\it decreases} -- qualitatively consistent with the positive 
correlation seen in the ISCCP IR data between cosmic ray rate and low clouds 
(\cite{svensmark97}; Marsh and Svensmark 2000; Pall\'{e} Bag\'{o} and Butler 
2000). Therefore the high clouds appear to become thicker but smaller in 
response to increasing cosmic ray flux, while for the low clouds the 
response is the opposite. 

\section{Discussion}
\label{discussion}

The observed variation of cloud optical thicknesses with cosmic-ray rate 
can be used to constrain microphysical models of the cloud condensation 
nuclei concentration $N_{\rm{CCN}}$ using (\ref{dtaureff}). Of crucial 
importance is the partial derivative $\partial N_{\rm{CCN}}/\partial 
q$, which determines the sign of the change in opacity with cosmic-ray 
rate. Recently Yu (2002) calculated $N_{\rm{CCN}}$ as a function of 
altitude and ionization rate using an ion-mediated nucleation code. 
Given this model and the vertical profiles of sulfuric acid vapor 
concentration, ionization rate, temperature, relative humidity, pressure, 
and surface area of pre-existing particles assumed therein (Yu 2002), the 
value of $\partial N_{\rm{CCN}}/\partial q$ peaks at values of $q_{peak}=12$, 
$8$, and $4$ ion pairs cm$^{-3}$ for low, mid-level, and high clouds, 
respectively, such that $\partial N_{\rm{CCN}}/\partial q>0$ for 
$q<q_{peak}$ and $\partial N_{\rm{CCN}}/\partial q<0$ for $q>q_{peak}$. 
Using the cosmic-ray ionization rates found by Neher (1961,1967) 
interpolated to geomagnetic latitude $40^{\circ}$, we find typical 
ionization rates of, respectively, $q\sim 3$, $8$, and $23$ ion pairs 
cm$^{-3}$ for the low, mid-level, and high ISCCP clouds. Therefore from 
(\ref{dtaureff}) we would expect a positive or zero correlation between 
opacity and cosmic-ray rate only for low clouds, and negative correlations 
for higher clouds for this model. We observe just the opposite, but the 
precision of the temporal model fits to the ISCCP data is not sufficient 
for us to rule out the Yu (2002) model based on the 
data. 

All three of the time-varying parameters in our temporal model show inverse 
correlations between mean visible cloud opacity and amount, suggesting a 
common origin for this behavior. These inverse correlations are illustrated 
in Figure~9. These are probably not artifacts of the averaging process because 
the quantities in Figure~9 have been normalized by their constant model terms 
in their temporal fits. One possible explanation for the inverse opacity-amount 
correlation is via a feedback mechanism. For the case of positive opacity 
variations, an increase in mean cloud opacity and albedo would result in 
increased energy loss to space and eventually less surface heating and 
subsequent water evaporation. Hence clouds would tend to be smaller and have 
smaller areas than they would otherwise. Conversely, for negative opacity 
variations clouds would tend to be larger. Global climate simulations 
(\cite{chen96}) indicate that global cloud albedo-increasing perturbations 
-- similar to the changes induced by cosmic rays -- decrease the global 
transport of moisture from the tropics, which then could conceivably produce 
fewer or smaller global clouds on average by this mechanism. Dynamical 
simulations of the response of global cloudiness to synoptic changes in 
the opacity are needed to investigate this.

\section{Summary}
\label{summary}

Here we consider a model in which Galactic cosmic rays alter the 
optical properties of clouds by changing the number of available 
cloud condensation nuclei. The main observational consequence of 
our model is a change in mean cloud opacity, with a secondary 
effect being a change in infrared emittance for optically thin 
clouds due to the relationship between cloud emissivity and opacity. 
We use the global ISCCP cloud database to search for variations in 
cloud properties due to cosmic rays, and after subtracting the background 
signals in the data due to Mt. Pinatubo and ENSOs, we find systematic 
variations in both opacity and cloud amount associated with changes 
in the cosmic-ray rate. The fractional variation in opacity attains 
a maximum positive value for high clouds and decreases with height, 
becoming negative or zero for low clouds. The fractional variation 
of the cloud amounts with cosmic-ray rate, however, show the opposite 
trend -- increasing from a negative correlation at high altitudes to 
a positive correlation at low altitudes, which is consistent with the 
positive correlation between global low clouds clouds and cosmic-ray 
rate seen in the infrared (\cite{svensmark97}; Marsh and Svensmark 2000; 
Pall\'{e} Bag\'{o} and Butler 2000)

Clearly more work is needed to model the opacity and cloud amount 
variations seen in the ISCCP data. Using our simple temporal model 
and perturbative approach, we have outlined a framework on which 
the variations in the data due to cosmic rays can be isolated 
and compared to  model predictions. As the time span of the 
ISCCP data increases in length, more complicated models with 
additional components and nonlinear dependencies can be used, 
and the analysis can then be more robust. The ISCCP data requires the 
culling together and normalizing of many disparate satellite datasets 
(\cite{rossow99}), and although this approach is necessary at the 
present time it is not ideal. One complement to the ISCCP global 
cloud data would be provided by the NASA deep space mission {\it Triana}, 
which would be able to retrieve cloud optical thicknesses simultaneously 
over the entire sunlit Earth from the L1 Lagrangian point between the 
Earth and the Sun. Continuous deep space observing of Earth's clouds 
would be ideal for detecting not only the solar cycle variations seen 
here but also the shorter duration but possibly more frequent variations 
in global cloud cover associated with Forbush decreases of Galactic 
cosmic rays and high energy solar proton events from the Sun. 

\acknowledgments

We thank the AVANTI article service of the Scripps Institution of 
Oceanography Library, and acknowledge the use of cosmic-ray data from 
the University of Chicago (National Science Foundation grant ATM-9912341) 
and Southern Oscillation Index data from the Australian Bureau of 
Meteorology. We also would like to thank the anonymous referees for 
very helpful comments.

{}

\newpage

\begin{planotable}{lcccccr}
\tablewidth{0pt}
\tablecaption{Temporal Fits to ISCCP Mean Cloud Opacity ($\bar{\tau}$) 
and Cloud Amount ($A$)\tablenotemark{1}}
\tablehead{\colhead{Par.} & \colhead{Lat.} & \colhead{Alt.} & 
\colhead{$b_{0}$} & \colhead{$b_{1}$} & \colhead{$b_{2}$} & \colhead{$b_{3}$}\\
\colhead{} & \colhead{} & \colhead{} & \colhead{[Const.]} & \colhead{[SOI]} & 
\colhead{[Pinatubo]} & \colhead{[CR]}}
\startdata
$\bar{\tau}$ & $<40^{\circ}$ & High & $13.31\pm 0.23$ & $7.6\pm 
1.7$ & $-32.2\pm 1.6$ & $9.4\pm 1.6$\nl
$\bar{\tau}$ & $<40^{\circ}$ & Mid-level & $10.45\pm 0.32$ & 
$7.9\pm 2.9$ & $-28.3\pm 2.7$ & $5.8\pm 2.9$\nl
$\bar{\tau}$ & $<40^{\circ}$ & Low & $3.78\pm 0.14$ & $-1.9\pm
3.1$ & $-4.2\pm 3.6$ & $-2.5\pm 3.4$\nl
$\bar{\tau}$ & $>40^{\circ}$ & High & $12.00\pm 0.50$ & $16.6\pm 
4.0$ & $-6.9\pm 3.6$ & $10.9\pm 3.5$\nl
$\bar{\tau}$ & $>40^{\circ}$ & Mid-level & $6.81\pm 0.23$ & $12.4
\pm 3.0$ & $-9.7\pm 3.0$ & $1.8\pm 3.2$\nl
$\bar{\tau}$ & $>40^{\circ}$ & Low & $5.01\pm 0.14$ & $1.8\pm 
2.3$ & $-7.9\pm 2.6$ & $-6.7\pm 2.6$\nl
\tableline
$A$ & $<40^{\circ}$ & High & $13.84\pm 0.52$ & $-4.8\pm 0.32$ & $61.6\pm 
5.8$ & $-9.3\pm 3.6$\nl 
$A$ & $<40^{\circ}$ & Mid-level & $5.71\pm 0.16$ & $2.0\pm 2.5$ & $58.1
\pm 4.2$ & $-0.1\pm 2.7$\nl
$A$ & $<40^{\circ}$ & Low & $45.5\pm 1.6$ & $6.9\pm 3.3$ & $13.6\pm 4.0$ 
& $0.8\pm 3.5$\nl
$A$ & $>40^{\circ}$ & High & $15.90\pm 0.58$ & $-12.3\pm 3.3$ & $7.9\pm 
4.0$ & $-9.9\pm 3.3$\nl
$A$ & $>40^{\circ}$ & Mid-level & $23.61\pm 0.84$ & $-4.6\pm 3.0$ & 
$8.3\pm 3.8$ & $-3.1\pm 3.4$\nl
$A$ & $>40^{\circ}$ & Low & $38.7\pm 1.0$ & $7.1\pm 2.4$ & $10.1\pm 
2.9$ & $6.5\pm 2.6$\nl
\tablenotetext{1}{Except for the amplitude of the constant term $b_{0}$, 
the model amplitudes for the SOI, Mt. Pinatubo, and cosmic-ray terms 
($b_{1}$-$b_{3}$, respectively) are normalized to $b_{0}$ and expressed 
as a percentage. The mean cloud opacity $\bar{\tau}$ and total mean amount 
fraction $A$ are from the ISCCP visible band data.}
\end{planotable}

\clearpage

\begin{figure}
\caption{~Cartoon illustrating two limiting scenarios for the effect
of the Galactic cosmic rays (GCRs) on cloud optical properties,
assuming that the varying ionizing cosmic-ray flux causes changes in 
the number of cloud condensation nuclei (CCN) through ion-mediated 
nucleation. In the first case we assume that the nucleation of cloud 
droplets is limited by the available amount of water in the supersaturated 
air. Therefore, as illustrated in the top panel, if an increase in the 
GCR ionization flux resulted in more cloud condensation nuclei (CCN) but 
no additional water condensation, the amount of water per droplet will be 
less and the effective radius $R_{eff}$ of the droplet distribution will 
be smaller. Alternately, as illustrated in the bottom panel, if the formation 
of cloud droplets is limited by the local availability of CCN and not 
condensible water, $R_{eff}$ can remained unchanged and additional CCN 
resulting from changes in cosmic-ray ionization would cause an increase 
in the amount of water extracted from the supersaturated air, so the cloud 
liquid water content would increase. The opposite trends hold for cases 
where the number of CCN is decreased by variations in the cosmic-ray flux.}
\end{figure}

\begin{figure} 
\caption{~The fractional change in the albedo of a $1$ km thick cloud 
expected from a 10\% increase in the number of cloud droplets due to 
changes in the cosmic-ray flux, shown for the case of variable cloud 
water content LWC (top) and for variable droplet radius $R_{eff}$ (bottom). 
The solid contours denote the change in albedo, and the dotted contours are 
for the optical thickness.}
\end{figure}

\begin{figure}
\caption{~The fractional change in the albedo, expected from a 
10\% increase in the number of cloud droplets from variations in 
the cosmic-ray rate, plotted as a function of cloud optical thickness 
for three different cloud geometrical thicknesses. The open symbols 
denote changes in cloud LWC and the filled symbols changes in $R_{eff}$.}
\end{figure}

\begin{figure}
\caption{~The mean cloud $0.6$ $\mu$m optical thickness from the ISCCP
database for all clouds in the low latitude band $|\phi|<40^{\circ}$, 
with the cosmic-ray rate from the Climax neutron monitor. The high, 
mid-level, and low clouds refer to cloud top pressures of $P<440$ mb, 
$440<P<680$ mb, and $P>680$ mb, respectively, and the shaded interval 
refers to cloud data affected significantly by the eruption of Mt. 
Pinatubo in June 1991. The $1\sigma$ error bars on the ISCCP data 
were calculated from the sample variance of the data.}
\end{figure}

\begin{figure}
\caption{~Same as Figure~4, but for all high latitude clouds with  
$|\phi|>40^{\circ}$.}
\end{figure}

\begin{figure}
\caption{~The mean cloud $0.6$ $\mu$m amount fractions from the the 
ISCCP database for all clouds in the low latitude band $|\phi|<40^{\circ}$, 
with the cosmic-ray rate from the Climax neutron monitor. The $1\sigma$ 
error bars on the ISCCP data were calculated from the sample variance of 
the data.}
\end{figure}

\begin{figure}
\caption{~Same as Figure~6, but for all high latitude clouds with  
$|\phi|>40^{\circ}$.}
\end{figure}

\begin{figure}
\caption{~Basis vectors used in the temporal model of ISCCP visible 
opacity and amount fraction variations. The vectors $X_{0}$, $X_{1}$, 
$X_{2}$, and $X_{3}$ represent the constant level term and variations 
due to ENSO, the eruption of Mt. Pinatubo, and cosmic rays, respectively.}
\end{figure}

\begin{figure}
\caption{~Fractional change in visible cloud amount versus fractional 
change in visible opacity, from the fit of the temporal model to the 
ISCCP data. The points have been normalized by their respective constant 
term values in the temporal model, and the error bars have been omitted. 
In all cases there is an inverse correlation between the two parameters.}
\end{figure}

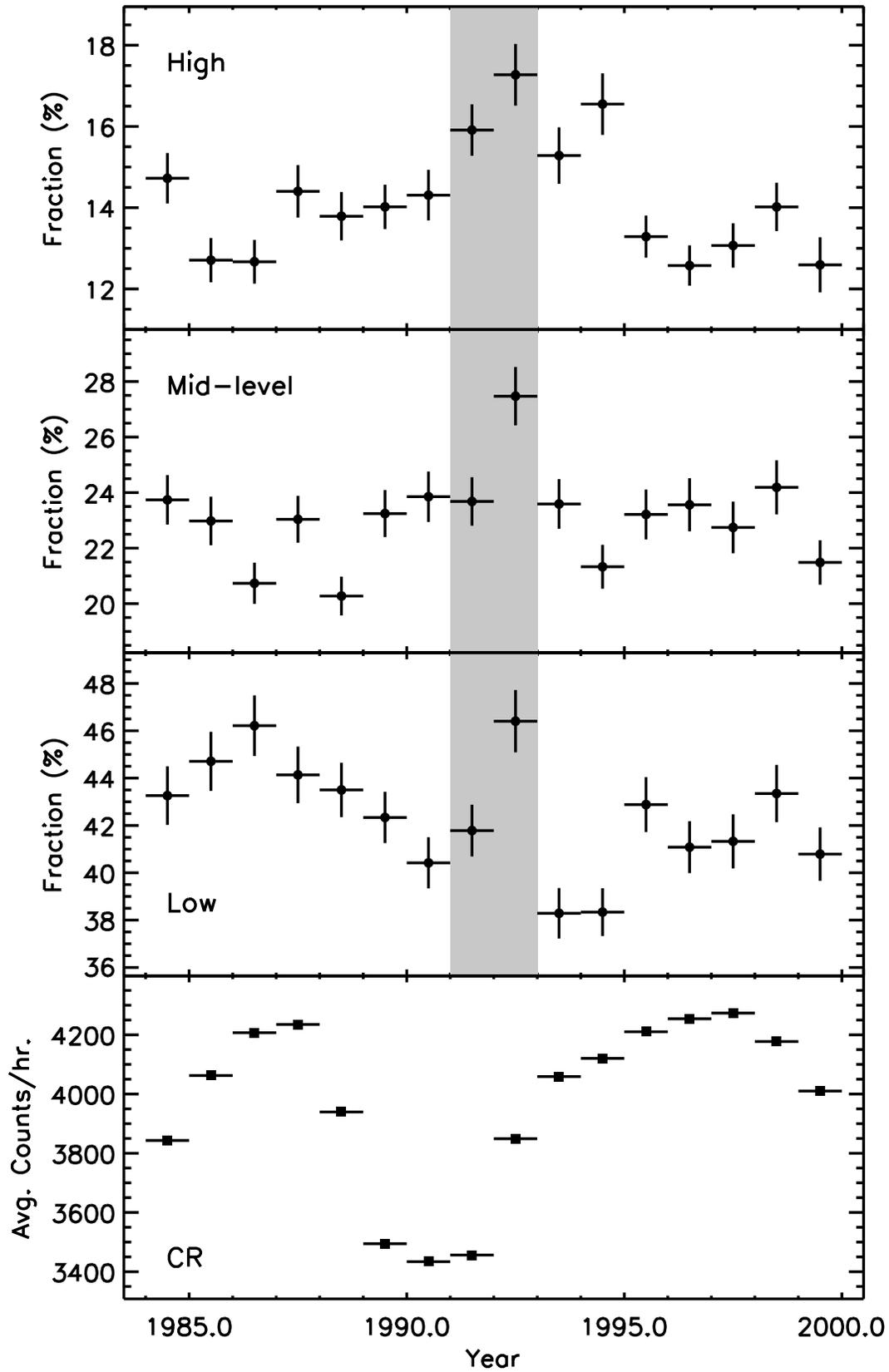
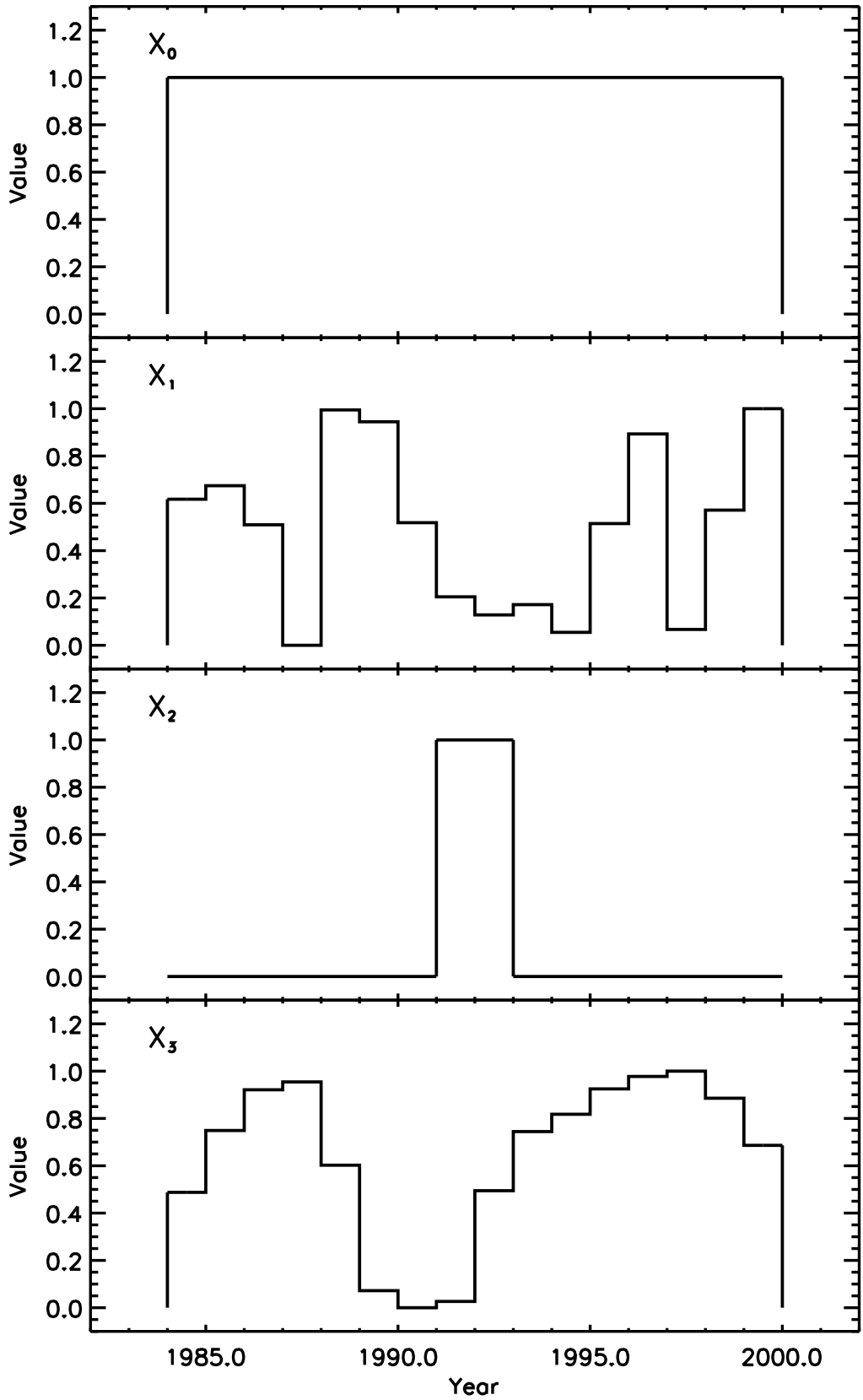
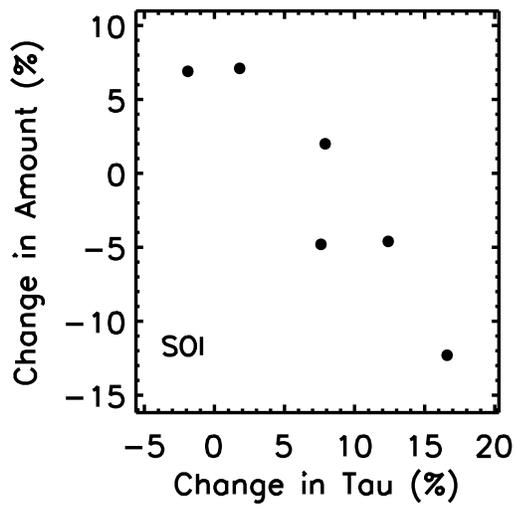
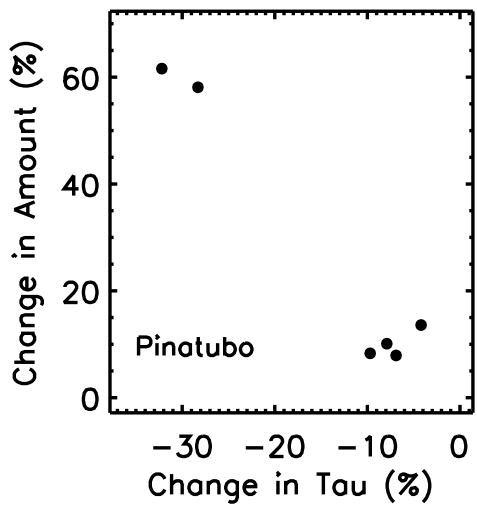
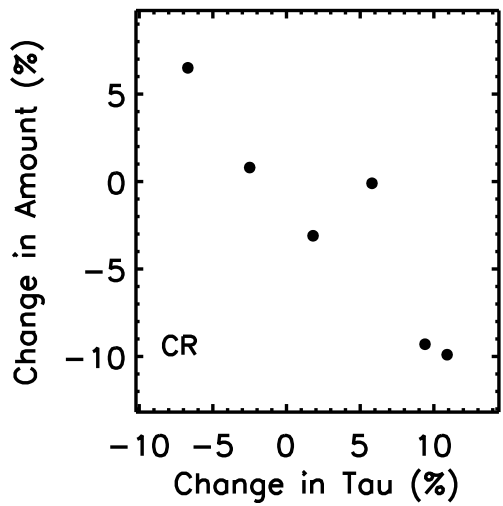
\end{document}